\documentclass[a4paper,11pt]{article}
\usepackage{pos}
\usepackage{tikz-feynman}
\usepackage{tikz}
\usepackage{siunitx}
\renewcommand{\logo}{\relax}
\title{Teaming up MET plus jet with Drell-Yan in the SMEFT}

\author[a]{Gudrun~Hiller}
\author[a]{Lara~Nollen}
\author*[a]{Daniel Wendler}
\affiliation[a]{TU Dortmund University\\ Department of Physics, Otto-Hahn-Str.4, D-44221 Dortmund, Germany}

\emailAdd{ghiller@physik.uni-dortmund.de}
\emailAdd{lara.nollen@tu-dortmund.de}
\emailAdd{daniel.wendler@tu-dortmund.de}

\abstract{The Standard Model Effective Field Theory (SMEFT) is a widely utilized framework for exploring new physics effects in a model-independent manner.
In previous studies, Drell-Yan collider data has emerged as a promising signature due to its energy enhancement relative to Standard Model predictions.
We present recent works, extending this approach by also considering the "missing energy + jet" signature, which can constrain related dineutrino couplings and similarly benefits from energy enhancement.
The combination of these observables allows for constraining a broader selection of operators and helps resolve flat directions in a global analysis.
Overall, the bounds probe the multi-TeV range, with the strongest reaching up to $10\; \text{TeV}$ for four-fermion interactions and $7 \;\text{TeV}$ for gluonic dipole interactions.
Furthermore, we find that low energy flavor observables improve limits by up to a factor of three for dipole operators.
We also estimate sensitivities to new physics at future hadron colliders including the $\sqrt{s} = 27 \;\text{TeV}$ HE-LHC and the $\sqrt{s} = 100 \; \text{TeV}$ FCC-hh.}

\FullConference{9th Symposium on Prospects in the Physics of Discrete Symmetries (DISCRETE2024)\\
 2–6 Dec 2024\\
Ljubljana, Slovenia\\}

\begin{document}
\newcommand{\emiss}{E_T^{\text{miss}}}

\maketitle
\section{Standard Model Effective Field Theory}
The Standard Model Effective Field Theory (SMEFT) is a widely used framework in the search for heavy new physics (NP) beyond the electroweak scale.
The corresponding Lagrangian is constructed out of the standard model (SM) degrees of freedom, under the assumption of $SU(3)_C \times SU(2)_L \times U(1)_Y$ invariance and a series of higher-dimensional operators $O_i^{(d)}$ with a corresponding Wilson coefficient (WC) $C_i^{(d)}$ and suppressed by a factor $\Lambda^{d-4}$, where $\Lambda$ is the NP scale.
Therefore, the Lagrangian can be written as 
\begin{equation}
  \mathcal{L}_{\text{SMEFT}} = \mathcal{L}_{\text{SM}} + \sum_d^{\infty} \sum_i \frac{C_i^{(d)}}{\Lambda^{d-4}} O_i^{(d)},
\end{equation}
where $\mathcal{L}_{\text{SM}}$ denotes the SM Lagrangian.
In this work we are focusing on terms at $d = 6$ and we adopt the Warsaw basis \cite{Grzadkowski:2010es}.
\section{Collider observables}
\subsection{Cross sections and energy enhancement}
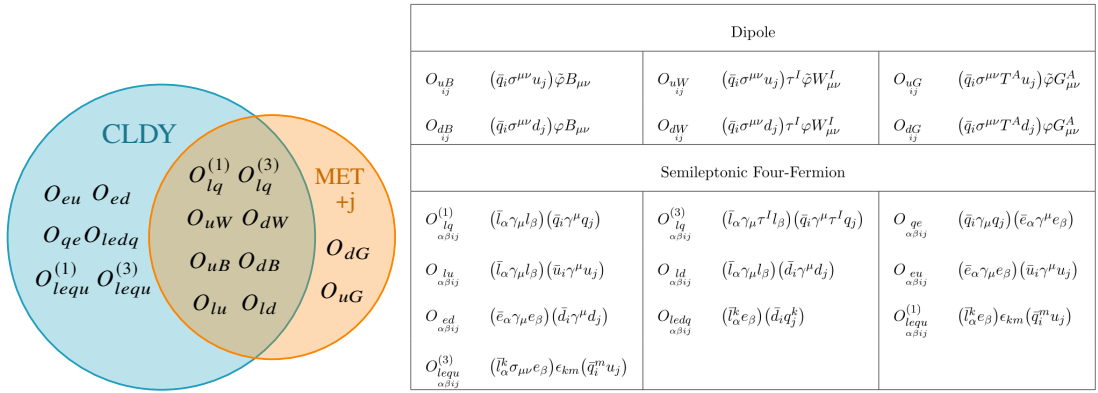
\begin{figure}[ht]
  \centering
  \resizebox{0.38\textwidth}{!}{\xdefinecolor{dRed}{RGB}{153, 0, 0}
\xdefinecolor{dOrange}{RGB}{251, 133, 0}
\xdefinecolor{dBlue}{RGB}{33, 158, 188}
\xdefinecolor{dOrangeDarker}{RGB}{201, 106, 0}
\xdefinecolor{dBlueDarker}{RGB}{25, 119, 141}
\xdefinecolor{dGrey}{RGB}{100, 100, 100}
\xdefinecolor{dBlack}{RGB}{0, 0, 0}

\begin{tikzpicture}
    \draw[color=dBlue, fill=dBlue, fill opacity=0.3, line width=0.3mm] (0,0) circle (2.5cm);
    \draw[color=dOrange, fill=dOrange, fill opacity=0.3, line width=0.3mm] (1.8,0) circle (2.0cm);

    \node[text=dBlueDarker, scale=1.2] at (-0.35,1.7) {CLDY};
    \node[text=dOrangeDarker] at (2.9,1.0) {MET};
    \node[text=dOrangeDarker] at (3,0.57) {+j};

    \node[text=] at (3.05,-0.2) {$O_{dG}$};
    \node[text=dBlack] at (2.95,-0.9) {$O_{uG}$};

    \node[text=dBlack] at (0.8,1.0) {$O_{lq}^{(1)}$};
    \node[text=dBlack] at (1.6,1.0) {$O_{lq}^{(3)}$};
    \node[text=dBlack] at (0.8,0.3) {$O_{uW}$};
    \node[text=dBlack] at (1.7,0.3) {$O_{dW}$};
    \node[text=dBlack] at (0.8,-0.4) {$O_{uB}$};
    \node[text=dBlack] at (1.6,-0.4) {$O_{dB}$};
    \node[text=dBlack] at (0.8,-1.1) {$O_{lu}$};
    \node[text=dBlack] at (1.6,-1.1) {$O_{ld}$};

    \node[text=dBlack] at (-1.6,0.7) {$O_{eu}$};
    \node[text=dBlack] at (-0.8,0.7) {$O_{ed}$};
    \node[text=dBlack] at (-1.6,0.0) {$O_{qe}$};
    \node[text=dBlack] at (-0.8,0.0) {$O_{ledq}$};

    \node[text=dBlack] at (-1.6,-0.7) {$O_{lequ}^{(1)}$};
    \node[text=dBlack] at (-0.6,-0.7) {$O_{lequ}^{(3)}$};

\end{tikzpicture}}
  \includegraphics[width=0.6\textwidth]{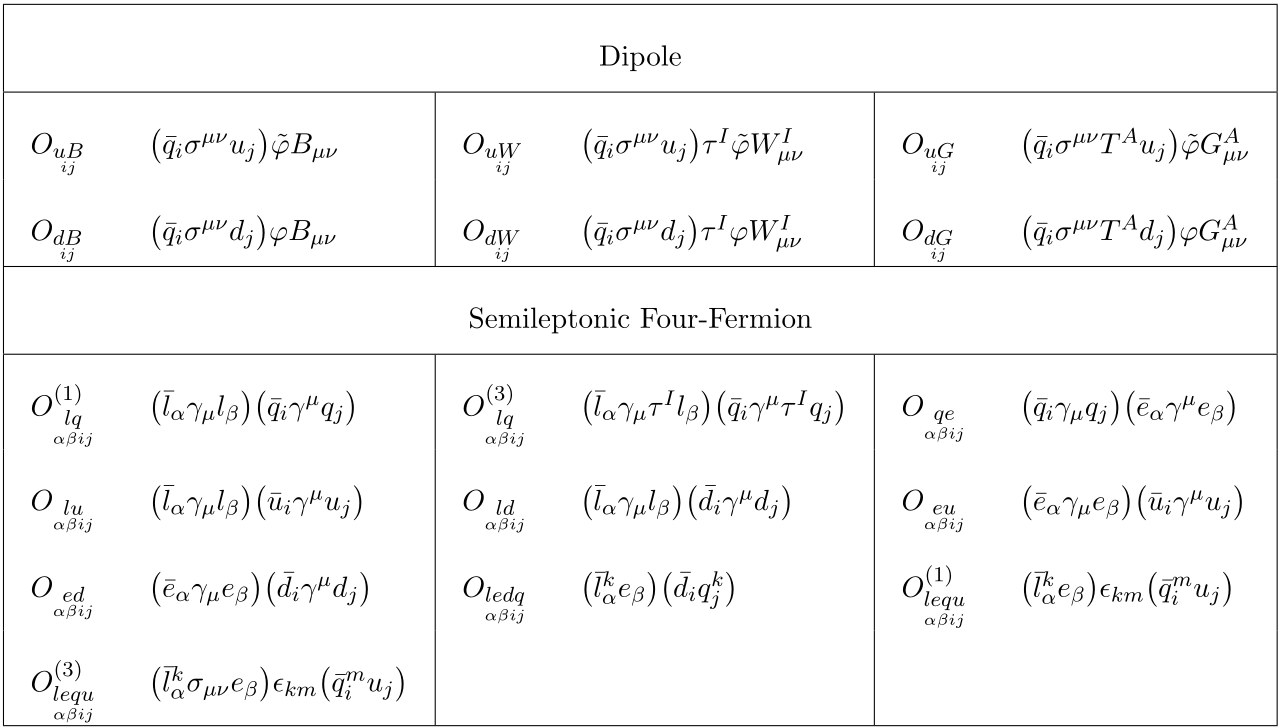}
  \caption{SMEFT operators considered in this work (right) and schematic depiction of the sensitivities (left) for the CLDY (blue) and MET+j (orange) observables.}
  \label{fig:operators}
\end{figure}

In the scope of our research we expand upon the previously examined charged lepton Drell-Yan (CLDY) process $pp \to \ell \ell^{(')}$ \cite{Grunwald:2023nli,Allwicher:2022mcg}, which probes couplings of quarks and charged leptons, by combining it with the signature of missing transverse energy and an energetic jet (MET+j)  $p p \to \nu \bar \nu +\text{j}$ \cite{Hiller:2024vtr}, which provides sensitivity to dineutrino couplings.
Within the SMEFT framework left handed charged leptons and neutrinos are part of a $SU(2)$-doublet, which implies correlations between operators and therefore synergies between the observables.  
The operators we consider and the overlap between within the observables is shown in Fig.~\ref{fig:operators}. 
Furthermore there are useful linear combinations defined by $C_{lq}^{\pm} = C^{(1)}_{lq} \pm C^{(3)}_{lq}$, $ C_{u(d)Z} = \pm s_W C_{u(d)W} + c_W C_{u(d)B}$ and $C_{u(d)\gamma} = \pm c_W C_{u(d)W} - s_W C_{u(d)B}$, where $s_W = \sin \theta_W, c_W = \cos \theta_W$ with the weak mixing angle $\theta_W$. 
Our focus within the SMEFT is on three main classes of operators: four-fermion contact terms (4F), electroweak dipole operators (EW) modifying the $Z-$ and $\gamma$-vertices with quarks and gluonic dipole operators (G), which affect the gluon-quark vertex.
The parton-level cross sections for the CLDY process can be parameterized as 
\begin{equation}
  \begin{aligned}
  \hat \sigma\left( q_i \bar q_j \rightarrow \ell_{\alpha}\ell_{\beta}\right) &= \delta_{ij} \delta_{\alpha \beta}  \hat \sigma_{\text{SM}}
  +  C^{\ell \ell}_{\underset{\alpha\beta ij}{\text{4F},\text{lin}}}   \delta_{ij} \delta_{\alpha \beta}  \hat\sigma_{\text{4F}}^{\text{lin}} 
  + \left(C^{\ell \ell}_{\underset{\alpha\beta ij}{\text{4F},\text{quad}}}\right)^2 \hat\sigma_{\text{4F}}^{\text{quad}}
 + \delta_{\alpha \beta } \left(C^{\ell \ell}_{\underset{ij}{\text{EW}}}\right)^2 \hat \sigma_{\text{EW}} \,,
\end{aligned}
\label{eqn:xsec_ll}
\end{equation}
and the leading contribution for the MET+j process as 
\begin{equation}
  \begin{aligned}
    \frac{\mathrm{d}\hat \sigma\left( q_i g \rightarrow \nu \bar \nu q_j\right) }{\mathrm{d} E_T^{\text{miss}}} &= \frac{\mathrm{d}\hat \sigma_{SM} }{\mathrm{d} E_T^{\text{miss}}} \delta_{ij} 
    +C^{\nu \nu}_{\underset{ij}{\text{4F},\text{lin}}} \delta_{ij}  \frac{\mathrm{d}\hat \sigma_{4F}^{\text{lin}}}{\mathrm{d} E_T^{\text{miss}}} 
     +\left(C^{\nu \nu}_{\underset{ij}{\text{4F},\text{quad}}}\right)^2\frac{\mathrm{d}\hat \sigma_{4F}^{\text{quad}} }{\mathrm{d} E_T^{\text{miss}}} \\
    &+\left(C^{{\nu \nu}}_{\underset{ij}{\text{EW}}}\right)^2 \frac{\mathrm{d}\hat \sigma_{EW} }{\mathrm{d} E_T^{\text{miss}}} +\left(C^{\nu \nu}_{\underset{ij}{\text{G}}}\right)^2 \frac{\mathrm{d}\hat \sigma_{G} }{\mathrm{d} E_T^{\text{miss}}} \,,
  \end{aligned}
  \label{eqn:xsec_nunu}
\end{equation}
where we introduce effective WCs $C^{\ell \ell (\nu \nu)}_{I}, \; I = \{ \text{4F}, \text{EW}, \text{G}\}$ which are directly probed in the process and consist of different linear combinations of SMEFT WCs in Fig.~\ref{fig:operators}.
The parameterizations show that linear term only appear for the four-fermion type operators, as these interfere with SM processes for diagonal quark and lepton flavor. 
Interference terms for dipole operators are proportional to quark masses and therefore strongly suppressed for the energy scales we consider.
One feature of the different classes of operators is their relative energy enhancement compared to the SM cross section, leading to increased sensitivities in the tails of both observables.
This is especially pronounced for the four-fermion and gluonic dipole operators, whose cross sections growth with the square of partonic center-of-mass-energy $\hat s^2$ relative to the SM cross section.
Whereas for EW dipole operators the cross section only growths with a single power of $\hat s$ relative to the SM cross section.\\
Furthermore, the effective WCs $C^{\ell \ell (\nu \nu)}_{I}$ probe different combination of SMEFT WCs, leading to the resolution of flat directions.
For the WCs $C^{(1)}_{lq},C^{(3)}_{lq}$ different linear combinations are probed, depending on which sector is considered.
For up-type quarks the combination $C^{+}_{lq} = C^{(1)}_{lq} + C^{(3)}_{lq}$ is probed by MET+j,
whereas $C^{-}_{lq} = C^{(1)}_{lq} - C^{(3)}_{lq}$ is probed by CLDY.
If we instead consider down-type quarks the $SU(2)$-flip leads to a different sign, i.e. $C^{-}_{lq} $ is probed by MET+j  and $C^{+}_{lq}$by CLDY.
As the photon does not couple to neutrinos in the SM, the  $C^{ \nu \nu}_{\text{EW}}$ only probes $C_{u(d)Z}$, whereas $C^{ \ell \ell}_{\text{EW}}$ is proportional to both $C_{u(d)Z}$ and $C_{u(d)\gamma}$. 
The combination of this allows to constrain the SMEFT parameters $C_{u(d)W},C_{u(d)B}$. \\
To obtain the full hadronic cross section, Eq.\eqref{eqn:xsec_ll} and  Eq.\eqref{eqn:xsec_nunu} have to be convoluted with the parton distribution functions (PDFs) of the proton, see for example \cite{Grunwald:2023nli,Hiller:2024vtr}.
For the CLDY process only the $q_i\bar q_j$ channel contributes, whereas for the MET+j process $ q_i\bar q_i$,$q_i g$ and $\bar q_i g$ contribute, which can be similarly parameterized as Eq.~\eqref{eqn:xsec_nunu}.
\subsection{Flavor rotations}
To diagonalize the Yukawa matrices, one must redefine the quark fields in terms of the mass eigenstate fields, which subsequently leads to the appearance of rotation matrices in the effective operators listed in Table 1. For most operators, the Wilson coefficients can then be redefined to uniquely absorb these rotation matrices.
However, this is not the case for four-fermion operators involving $SU(2)_L$ doublets.
As an example, we consider the operators $O^{(1)}_{lq}, O^{(3)}_{lq}$, with the explicit direction $C^{+}_{lq} \neq 0,C^{-}_{lq} = 0$.
The Lagrangian then reads 
\begin{equation}
  \begin{aligned}
    \label{eqn:Lagr}
    \mathcal{L}^{6} &\supset
    C^{+}_{\underset{\alpha \beta ij}{lq}} \Bigl( \bigl(\bar u_i \gamma^{\mu}P_L u_j \bigr) \bigl(\bar \nu_{\alpha} \gamma_{\mu} \nu_{\beta} \bigr) + \bigl( \bar d_i \gamma^{\mu}P_L d_j \bigr) \bigl( \bar \ell_{\alpha} \gamma_{\mu} \ell_{\beta} \bigr) \Bigr) \\
    &= C^{+}_{\underset{\alpha \beta ij}{lq}} \left( U^{*u}_{\underset{ik}{L}} U^{u}_{\underset{jm}{L}} \left( \bar u^{\prime}_k \gamma^{\mu} P_L u^{\prime}_m \right) \left(\bar \nu_{\alpha} \gamma_{\mu} \nu_{\beta}\right) + U^{*d}_{\underset{ik}{L}} U^{d}_{\underset{jm}{L}} \left(\bar d^{\prime}_k \gamma^{\mu} P_L d^{\prime}_m \right) \left(\bar \ell_{\alpha} \gamma_{\mu} \ell_{\beta} \right) \right) \,.
  \end{aligned}
\end{equation} 
where in the second line, we performed the quark field transformations according to $q = U^{q}_{L} q^{\prime}$ for left-chiral fields $q = u,d$.  
In Eq.~\eqref{eqn:Lagr} there are now two choices to absorb the quark rotation matrices, which are given by  
up(down)-alignment 
, $ \hat C^{+, \text{up(down)}}_{lq} = {U^{u(d)}_{L}}^{\dagger}   C^{+}_{lq} U^{u(d)}_{L}$ which leads to factors of the CKM matrix $V = U^{u\dagger}_L U_L^d$ appearing for the down(up)-sector. 
For the other direction $C^{-}_{lq}  $, as well as for $C_{qe}$ similar arguments apply and in the following up(down)-alignment will always imply that all relevant WCs $C^{(1)}_{lq},C^{(3)}_{lq},C_{qe}$ are redefined in the same way.
Importantly, this also implies that even if we focus on quark off-diagonal flavor combinations, through the rotations we will always induce diagonal contributions which interfere with the SM.  
Regarding the lepton rotation matrices, we always assume them to be aligned with the charged leptons, as the MET+j is not sensitive to lepton flavor.

\section{Fit and results}
\begin{figure}[h]
  \centering
  \includegraphics[width=0.98\textwidth]{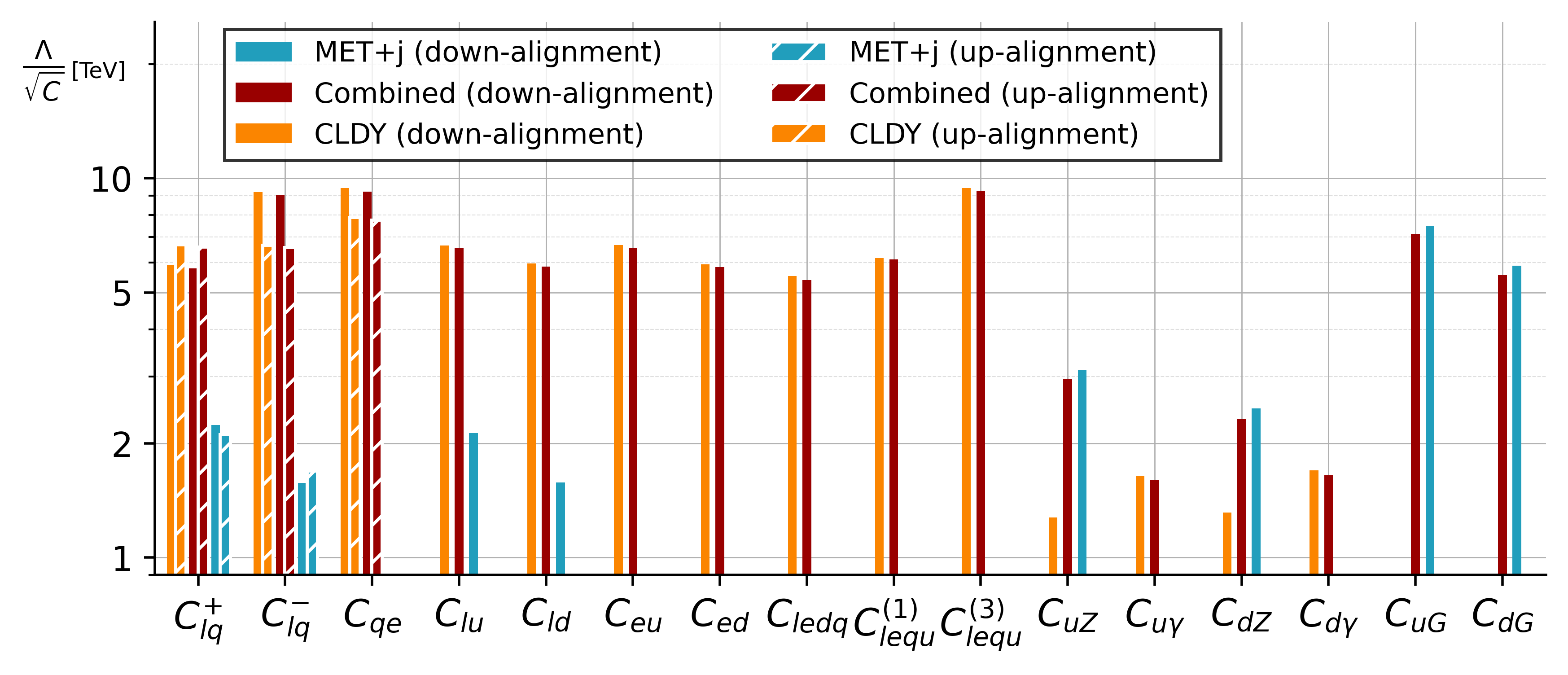}
  \includegraphics[width=0.48\textwidth]{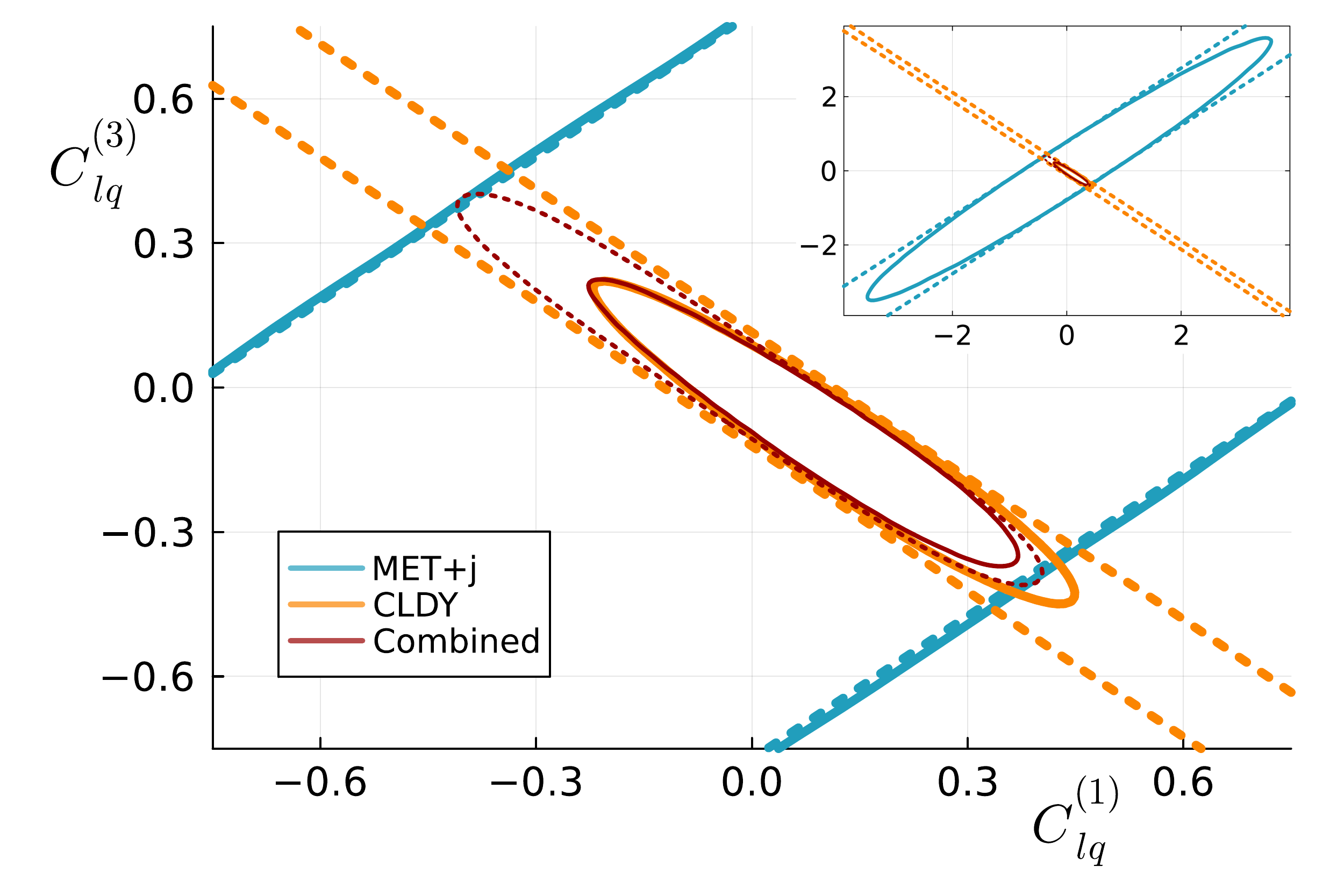}
  \includegraphics[width=0.48\textwidth]{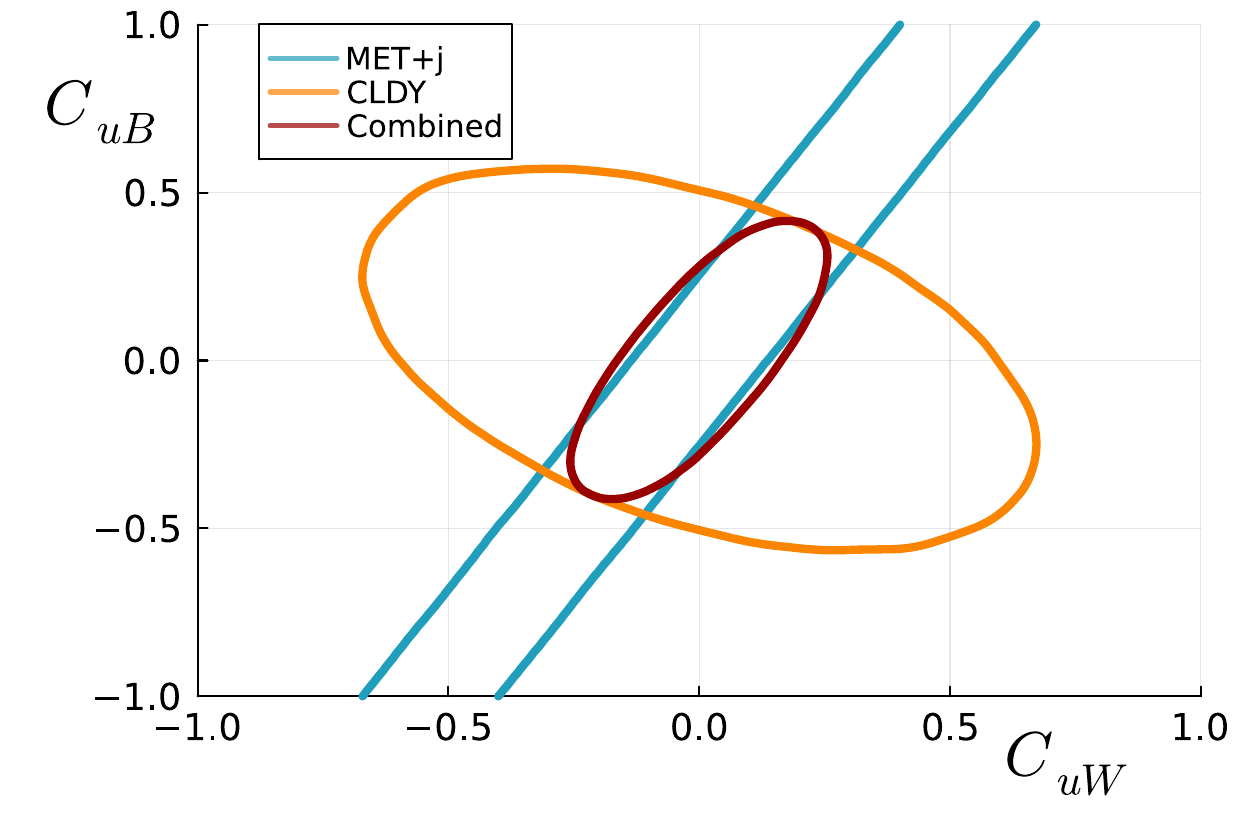}
  \caption{Results from the combined fit (red) and the individual fits of the CLDY (orange) and MET+j (blue) processes in the lepton flavor universal scenario.
  Shown are 95\% limits on $\Lambda/\sqrt{C}$ (top) for the quark indices $i,j=1,2$
  , the correlation between $C_{lq}^{(1)}$ and $C_{lq}^{(3)}$ for $i,j = 1,3$ (lower left)  and the correlation between $C_{uB}$ and $C_{uW}$ for $i,j = 1,2$ (lower right).
  Limits in down-alignment (up-alignment) are shown as solid (dashed) lines.  }
  \label{fig:results}
\end{figure}

We perform fits to the WCs in Fig.~\ref{fig:operators} as free parameters for fixed quark generation indices $i \neq j$ utilizing the EFT\textit{fitter} framework \cite{Castro:2016jjv} and the datasets obtained at the Large Hadron Collider (LHC) from Ref.~\cite{ATLAS:2020tre,CMS:2021ctt,ATLAS:2020zms} for CLDY and Ref.~\cite{ATLAS:2021kxv} for Met+j.
Furthermore, we adopt different scenarios for the lepton flavor of the four-fermion operators: Lepton-flavor-specific couplings, a lepton-flavor universal scenario and a lepton-flavor violating scenario.
Since the CLDY process distinguishes individual lepton flavors, the choice of scenario becomes more significant. For the MET+j process, the dineutrinos are not tagged and therefore are incoherently summed over. 
This implies that only the sum of squares of different WCs is constrained and by assuming the different scenarios, the bounds improve by fixed factors stemming from the combinatorics of the WCs within the sum.
In Fig. \ref{fig:results} we show the obtained $95 \%$ bounds on the NP scale $\Lambda/ \sqrt{C}$ for $i,j = 1,2$, as well as correlations between left handed four-fermion operators and EW dipole operators \cite{Hiller:2025hpf}.
Generally, bounds for $uc$-transitions are the strongest, as the PDFs are the largest and bounds on the other generations\cite{Hiller:2025hpf} are correlated through relative PDF factors.

\section{Impact of flavor data}
In general, flavor data constrains four-fermion operators several orders of magnitude more stringently, with the exception of WCs involving right handed taus, as these are kinematically forbidden for the first two generations and are not linked to dineutrino data through $SU(2)_L$.
We therefore focus on the synergies between our observables and flavor data for dipole type operators.
For this we utilize the weak effective theory (WET) framework, in which the relevant WET photon and gluon dipole operators are given by 
\begin{equation}
  \begin{aligned}
    &O_{\underset{ij}{7}}^{(\prime)} = \frac{e}{16\pi^2} m_i \left( \bar q^i_{L(R)} \sigma^{\mu\nu} q^j_{R(L)}\right) F_{\mu\nu} \,, \\
    &O_{\underset{ij}{8}}^{(\prime)} = \frac{g_s}{16\pi^2} m_i \left( \bar q^i_{L(R)} \sigma^{\mu\nu} T^A q^j_{R(L)}\right) G^A_{\mu\nu} \,,
  \end{aligned}
  \label{eqn:WET_dipoles}
\end{equation}
where $F_{\mu \nu}$ and $G^A_{\mu \nu}$ are the photon and gluon field strength tensors, respectively.
We neglect contributions from four-fermion WET operators, that arise by integrating out the $Z$-boson and obtain the combined constraints by running up to $\SI{10}{\TeV}$ and matching at the intermediate electroweak scale onto the SMEFT.
Generally, the combination with flavor improves the constraints by up to a factor 3 \cite{Hiller:2025hpf}, whereas for $uc$-transitions there is no improvement for $C_{uG}$, as the constrain on $C_8$ is relatively weak and the MET+j gives overall the strongest bound on this type of operator.

\section{Future collider sensitivities}
\begin{figure}[h]
  \includegraphics[width = 0.48\textwidth ]{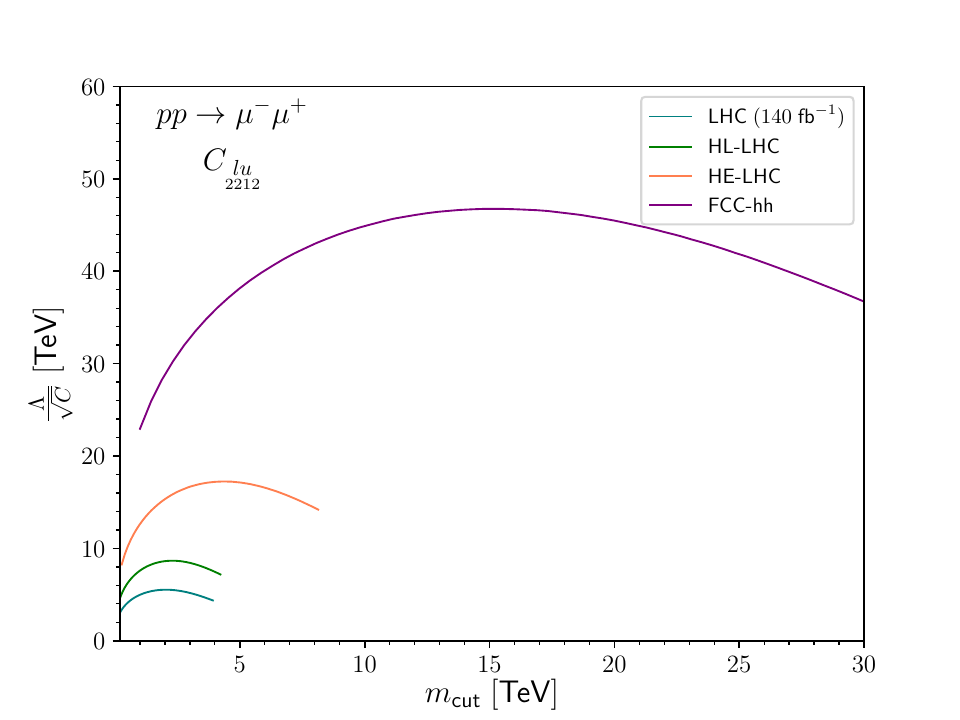}
  \includegraphics[width = 0.48\textwidth ]{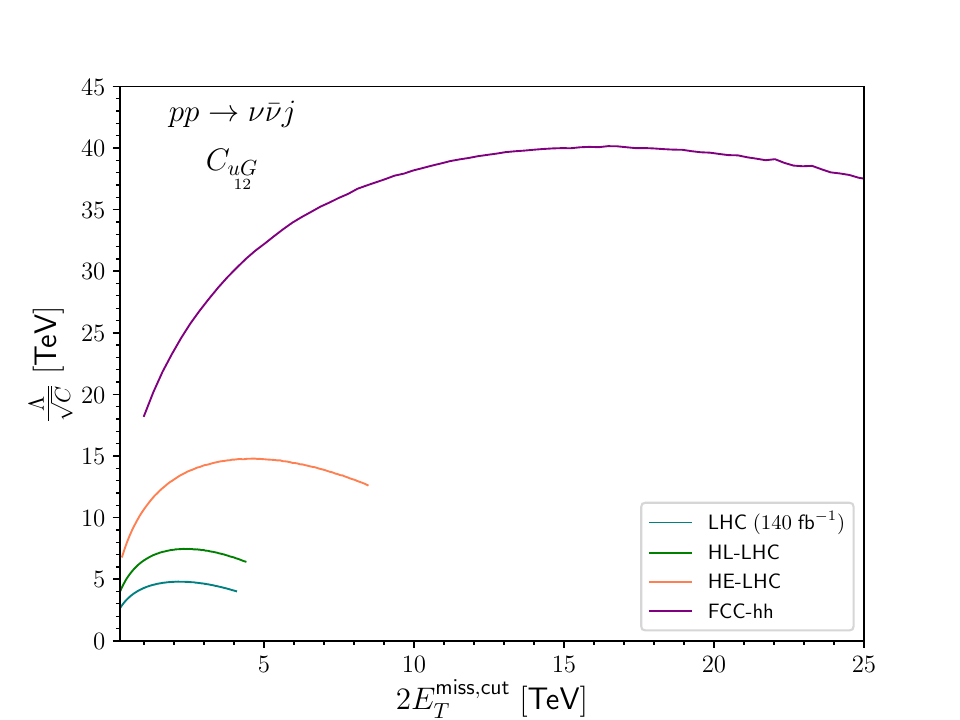}
  \caption{Estimated reach on the NP scale~$\Lambda$ as a function of the kinematic cut of the inclusive bin for the $m_{\mu \mu}$ observable (left) and the $\emiss$ observable (right) for a $uc$ transition. 
  The considered signal processes are the $O_{\underset{2212}{lu}}$ and the $O_{\underset{12}{uG}}$ for $i \neq j$ contributions, respectively.}
  \label{fig:FutCollider}
\end{figure}

To estimate the reach at future hadron colliders, we estimate the NP reach based on a inclusive one-bin analysis for the CLDY and MET+j observables.
The collider benchmarks we consider are the LHC ( $\sqrt{s} = \SI{13}{\TeV}, \mathcal{L} = \SI{0.14}{\atto\barn}$), the high-luminosity LHC (HL-LHC,$\sqrt{s} = \SI{14}{\TeV}, \mathcal{L} = \SI{3}{\atto\barn}$), high-energy LHC (HE-LHC,$\sqrt{s} = \SI{27}{\TeV}, \mathcal{L} = \SI{15}{\atto\barn}$)  and the Future Circular Collider (FCC-hh,$\sqrt{s} = \SI{100}{\TeV}, \mathcal{L} = \SI{20}{\atto\barn}$). 
For the CLDY process we consider the operator $O_{\underset{22 ij}{lu}}$ and for MET+j the operator $O_{\underset{ij}{uG}}$, each for $i\neq j$.
Furthermore, we assume that only one dominant background process contributes, which is given by $p p  \to  \mu^+ \mu^-$, through an intermediate $Z$-boson or photon, which is a very good approximation for the CLDY observable.
For the MET+j process the dominant background is given by $p p \to  \nu \bar \nu j$, with an intermediate $Z$-boson, which roughly only accounts for $70\%$ of the background and therefore is only a first order approximation.
As we examine one inclusive high energy bin, the lower edge of this bin is a priori not known. 
To estimate this lower edge at future colliders, we calculate the significance for different values of the cut and select the optimal value to enhance our sensitivity.
Through this, the analysis is sensitive to the convolution of different PDFs, depending on the initial state partons.
In Fig.~\ref{fig:FutCollider} we show the sensitivity to the NP scale $\Lambda / \sqrt{C}$ for CLDY and MET+j, as a function of the respective kinematic cut.
We observe that the sensitivity peaks for both processes and therefore the choice of the cut $m_{\text{cut}}$ or $E_T^{\text{miss},cut}$ plays an important role to obtain an optimal bound.
The optimal cut and the respective limit on $\Lambda$ for the FCC-hh are given by $m_{\text{cut}} \sim \SI{15}{\TeV}$ and $\Lambda  \sim \SI{47}{\TeV}$ for $C_{\underset{22 12}{lu}}$ and  
$E_T^{\text{miss,cut}} \sim \SI{8.2}{\TeV}$ and  $\Lambda  \sim \SI{40}{\TeV}$ for $C_{\underset{12}{uG}}$.
Results for the other flavors and collider benchmarks can be found in Ref~.\cite{Hiller:2024vtr}.
We also emphasize the point that in a future analysis the bins should be optimized for different initial state partons, as this impacts the sensitivities.
\section{Conclusions}
We present a SMEFT analysis of the combination of CLDY and MET+j observables at the LHC. 
We show that the combination of the observables leads to the resolution of flat direction in the SMEFT parameter space. 
Bounds probe the multi TeV range, with the strongest constraints for $uc$-transitions, reaching up to $\SI{10}{\TeV}$ for four-fermion operators and up to $\SI{7}{\TeV}$ for gluonic dipole operators.
Low energy data from flavor observables still play a crucial role, as the combination can improve the bound on EW dipole operator by up to a factor of three.
Furthermore, future colliders can improve the reach on the NP scale $\Lambda$ by a factor of 3 at the HE-LHC and up to a factor 8 at the FCC-hh. 
We also show that the flavor content of the considered operator can influence the binning of distributions and should be accounted for in future collider analyses.
\begin{acknowledgments}
  LN is supported by the doctoral scholarship program of the {\it Studienstiftung des Deutschen Volkes}.
  DW is very grateful to the organizers to be given the opportunity to present this work.
 \end{acknowledgments}

\end{document}